\documentclass[
    ,final            
  ]
  {aipproc}

\layoutstyle{6x9}

\begin{document}

\title{Exploring Neutron-Rich Oxygen Isotopes with MoNA}


\author{\setcounter{footnote}{1} N. Frank$^{1}$\footnotetext{Present address: Department of Physics, Concordia College, Moorhead, MN 56562}}{
address={National Superconducting Cyclotron Laboratory, \\ Michigan State University, East Lansing, MI 48824},
altaddress={Department of Physics \& Astronomy, Michigan State University, East Lansing, MI 48824}
}
\author{T. Baumann}{
address={National Superconducting Cyclotron Laboratory, \\ Michigan State University, East Lansing, MI 48824}
}
\author{D. Bazin}{
address={National Superconducting Cyclotron Laboratory, \\ Michigan State University, East Lansing, MI 48824}
}
\author{J. Brown}{
address={Department of Physics, Wabash College, Crawfordsville, IN 47933}
}
\author{P.A. DeYoung}{
address={Department of Physics, Hope College, Holland, MI 49423}
}
\author{\\ J.E. Finck}{
address={Department of Physics, Central Michigan University, Mt.\ Pleasant, MI 48859}
}
\author{A. Gade}{
address={National Superconducting Cyclotron Laboratory, \\ Michigan State University, East Lansing, MI 48824},
altaddress={Department of Physics \& Astronomy, Michigan State University, East Lansing, MI 48824}
}
\author{J. Hinnefeld}{
address={Department of Physics \& Astronomy, \\ Indiana University at South Bend, South Bend, IN 46634}
}
\author{R. Howes}{
address={Department of Physics, Marquette University, Milwaukee, WI 53201}
}
\author{\setcounter{footnote}{2} J.-L. Lecouey$^{2,}$\footnotetext{Present address: Laboratoire de Physique Corpusculaire, IN2P3, 14050 Caen, France}}{
address={National Superconducting Cyclotron Laboratory, \\ Michigan State University, East Lansing, MI 48824}
}
\author{\\ B. Luther}{
address={Department of Physics, Concordia College, Moorhead, MN 56562}
}
\author{\setcounter{footnote}{3} W.A. Peters$^{3}$\footnotetext{Present address: Department of Physics \& Astronomy, Rutgers, The State University of New Jersey, Piscataway, NJ 08854}}{
address={National Superconducting Cyclotron Laboratory, \\ Michigan State University, East Lansing, MI 48824},
altaddress={Department of Physics \& Astronomy, Michigan State University, East Lansing, MI 48824}
}
\author{\setcounter{footnote}{4} H. Scheit$^{4,}$\footnotetext{Present address: Nishina Center for Accelerator Based Science, RIKEN, Wako, Saitama 351-0198, Japan}}{
address={National Superconducting Cyclotron Laboratory, \\ Michigan State University, East Lansing, MI 48824}
}
\author{A. Schiller}{
address={National Superconducting Cyclotron Laboratory, \\ Michigan State University, East Lansing, MI 48824}
}
\author{\\ M. Thoennessen}{
address={National Superconducting Cyclotron Laboratory, \\ Michigan State University, East Lansing, MI 48824},
altaddress={Department of Physics \& Astronomy, Michigan State University, East Lansing, MI 48824}
}

\begin{abstract}

The Modular Neutron Array (MoNA) was used in conjunction with a large-gap dipole magnet (Sweeper) to measure neutron-unbound states in oxygen isotopes close to the neutron dripline. While no excited states were observed in $^{24}$O, a resonance at 45(2) keV above the neutron separation energy was observed in $^{23}$O.
\end{abstract}

\maketitle


\section{Introduction}

Measurements of excited states are important for the understanding of nuclear structure. Traditional $\gamma$-ray spectroscopy yields high-resolution data on the level structure of excited states. However, very neutron-rich nuclei approaching the neutron dripline have only a few, if any, bound excited states that can be studied with $\gamma$-ray spectroscopy. For these nuclei, unbound excited states can be explored using the method of neutron decay spectroscopy \cite{Dea87,Kry93,Tho99,Zin97,Che01}.

A collaboration of ten colleges and universities designed and constructed the large area Modular Neutron Array (MoNA) in order to study neutron-unbound states in neutron-rich nuclei \cite{Lut03,Bau05,Pet07,Whi07}. The construction of the array was a novel approach to involving undergraduate students in forefront nuclear physics research \cite{Whi07,How05,Fed05}. Undergraduates built the individual modules at their home institutions and helped assemble the array at the National Superconducting Cyclotron Laboratory (NSCL). 

\begin{figure}
  \includegraphics[height=.35\textheight]{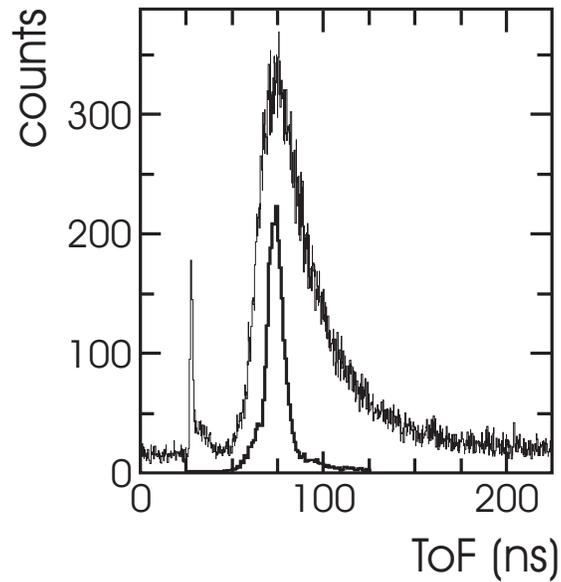}
  \caption{Time-of-Flight (ToF) spectrum measured in MoNA. The top curve was taken with a thick stopping target while the lower curve was taken with the production target and was gated by $^{22}$O fragments.}
  \label{tof}
\end{figure}

The goal of one of the first experiments with MoNA was the search for excited states in neutron-rich oxygen isotopes. The dripline nuclei $^{23}$O and $^{24}$O do not have any bound excited states \cite{Sta04} and the excitation energies of the first excited states in these nuclei are directly related to the size of the emerging $N$ = 14 and $N$ = 16 shell gaps \cite{Oza00,Thi00,Ots01}. 

\section{Experiment}

The excited states of the oxygen isotopes $^{23}$O and $^{24}$O were populated by a two-proton knockout reaction from a $^{26}$Ne secondary beam. Neutrons were detected with MoNA around zero degrees and the fragments were deflected by the Sweeper, a large-gap dipole magnet \cite{Bir05}, into a set of charged-particle detectors that provide position and angle information as well as particle identification \cite{Fra06}. Details about the production of the $^{26}$Ne beam and the experimental setup have been described elsewhere \cite{Pet07,Fra06,Fra07a,Sch07}.

The decay energies of excited fragments were reconstructed by the invariant mass method from the measured oxygen and neutron energies and the opening angle between the oxygen and the neutron. The angle and energy of the fragments were ion-optically reconstructed from the magnetic field of the Sweeper using the position and angle information recorded by the detectors located behind the Sweeper \cite{Fra07b}. The neutron energy was calculated from the time-of-flight (ToF) and the neutron angles were determined from the location of the first interaction within MoNA. 

Figure \ref{tof} shows the neutron ToF spectrum measured in MoNA. The upper solid line shows the ToF produced by stopping the beam in a thick target. A sharp peak corresponding to $\gamma$ rays produced in the interaction with the target is clearly visible. The known distance between the target and MoNA, the speed of light, and the location of this peak were used for the absolute ToF calibration. In addition to the prompt $\gamma$ peak neutrons at a ToF corresponding to the beam velocity are present. The neutron peak shows a long tail towards longer ToF originating from neutrons produced in more dissipative collisions. 

\begin{figure}
  \includegraphics[height=.4\textheight]{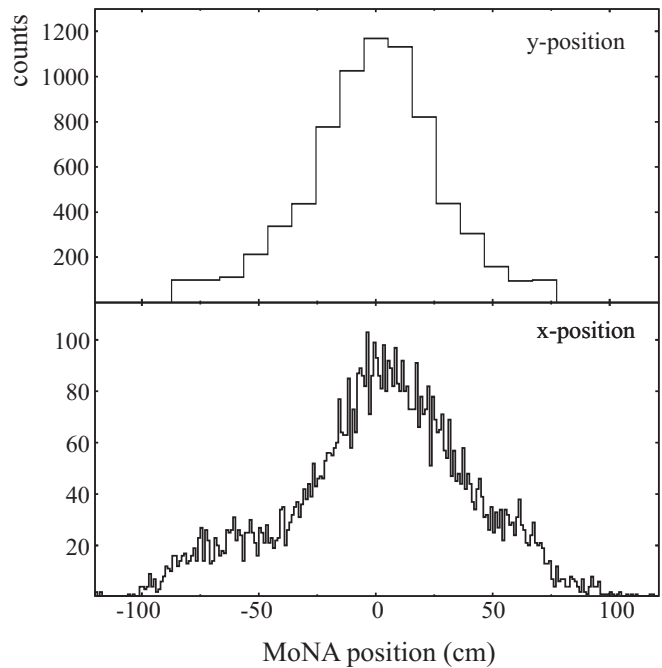}
  \caption{Y-position (top) and x-position (bottom) spectra for the first interaction within MoNA gated on $^{22}$O fragments.}
  \label{position}
\end{figure}

The lower curve was recorded with the production target in coincidence with $^{22}$O fragments measured in the charged-particle detectors behind the Sweeper. Only beam velocity neutrons corresponding to neutrons emitted from the intermediate $^{23}$O fragment are observed and the slower neutron tail is not present. A few remaining $\gamma$ rays were eliminated by the requirement that the first interaction of a neutron fragment coincidence had to occur after 40 ns.  

The angle of the neutrons was determined from the position of the first neutron interaction within MoNA. Figure \ref{position} shows the vertical (top) and horizontal (bottom) position distribution of neutrons gated on $^{22}$O. MoNA is 160 cm tall and 200 cm wide. In the y-direction the position resolution is determined by the 10 cm width of the individual detectors. The position in the x-direction is calculated from the time difference between the anode signals of the photomultiplier tubes mounted at the end of each detector. The strong angular distribution around zero degrees already indicates the presence of a low-lying resonance state in $^{23}$O. 

\section{Results}

\begin{figure}
  \includegraphics[height=.5\textheight]{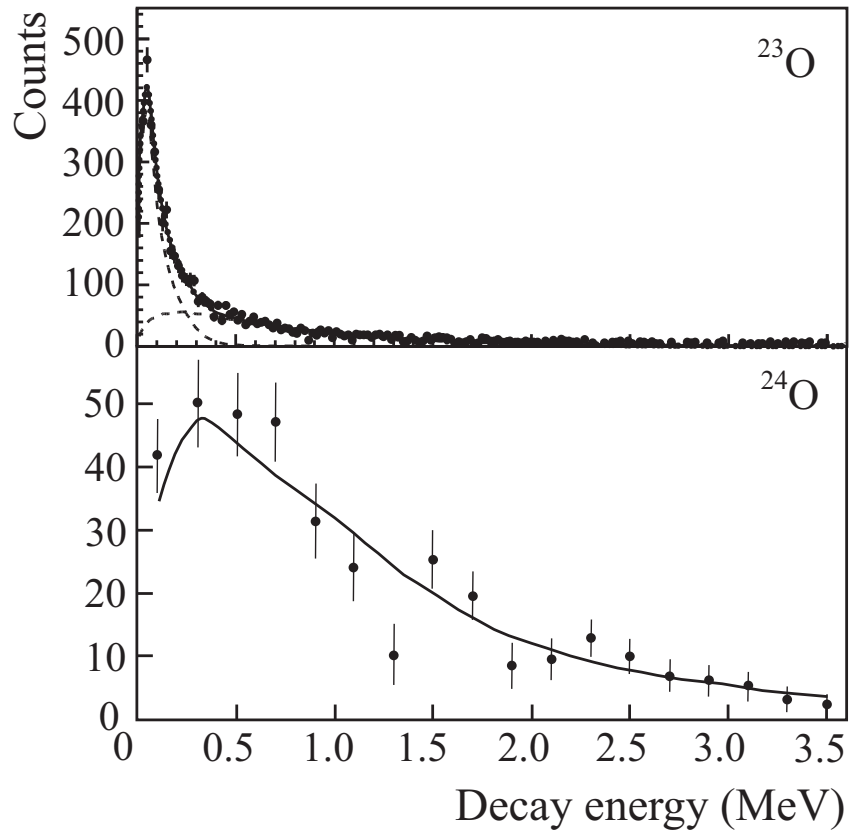}
  \caption{Decay energy spectra of $^{23}$O (top) and $^{24}$O (bottom).}
  \label{spec}
\end{figure}

Figure \ref{spec} shows the reconstructed decay energy spectrum of $^{23}$O (top) and $^{24}$O (bottom). The $^{24}$O spectrum can be described with a non-resonant background distribution shown by the solid line. The fit (solid line) to the $^{23}$O includes a sharp resonance at 45(2) keV in addition to background contribution (dashed lines). The absence of any resonances in $^{24}$O and the observation of only one resonance in $^{23}$O can be explained by describing the two-proton knockout as a direct reaction. Further details about the interpretation of the data are given in references \cite{Fra07a,Sch07,Fra07c}. 

\begin{figure}
  \includegraphics[height=.3\textheight]{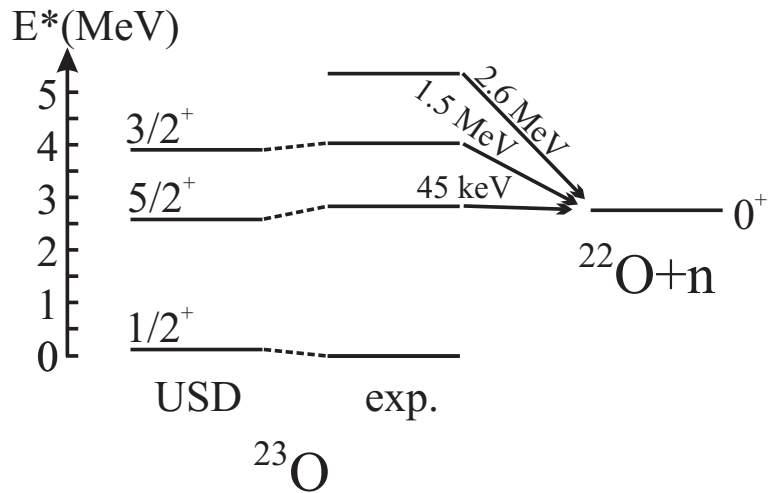}
  \caption{Level diagram of $^{23}$O. USD shell model calculations \cite{Bro06} are compared to the results from the present experiment (first excited state) and reference \cite{Ele07} (second and third excited states).}
  \label{level}
\end{figure}

The level diagram for $^{23}$O shown in Figure \ref{level} summarizes the current knowledge of this neutron-rich isotope. The present data and the results from reference \cite{Ele07} are compared to shell model calculations using the USDA interaction \cite{Bro06}. The third excited state has been interpreted as an intruder state which is not included in the USDA model space. The good agreement of the positive parity states confirms the large separation of the $d_{5/2}$, $s_{1/2}$ and $d_{3/2}$ orbitals.

\section{Conclusion}

The Modular Neutron Array was successfully used to observe the first excited state of $^{23}$O. In combination with the large-gap sweeper magnet, the setup can be used to measure many excited states that are not accessible with other techniques \cite{Fra07a}. In addition, the ground-state properties of nuclei beyond the dripline can be explored. In a first experiment the mass and decay width of $^{25}$O has been measured recently \cite{Hof07}.


\begin{theacknowledgments}
We would like to thank the members of the MoNA collaboration G. Christian, C. Hoffman, K.L. Jones, K.W. Kemper, P. Pancella, G. Peaslee, W. Rogers, S. Tabor, and about 50 undergraduate students for their contributions to this work. We would like to thank R.A. Kryger, C. Simenel, J.R. Terry, and K. Yoneda for their valuable help during the experiment. Financial support from the National Science Foundation under grant numbers PHY-01-102533, PHY-03-54920, PHY-05-55366 PHY-05-55445, and PHY-06-06007 is gratefully acknowledged. J.E.F. acknowledges support from the Research Excellence Fund of Michigan.
\end{theacknowledgments}



\bibliographystyle{aipproc}   

\bibliography{sample}

\IfFileExists{\jobname.bbl}{}
 {\typeout{}
  \typeout{******************************************}
  \typeout{** Please run "bibtex \jobname" to optain}
  \typeout{** the bibliography and then re-run LaTeX}
  \typeout{** twice to fix the references!}
  \typeout{******************************************}
  \typeout{}
 }


\end{document}